\documentclass[twocolumn,showpacs,preprintnumbers,prl,amsmath,amssymb]{revtex4}
\usepackage{graphicx}% Include figure files
\usepackage{dcolumn}% Align table columns on decimal point
\usepackage{bm}% bold math
\usepackage{amsmath}

\begin{document}

\title{Bolometric Detection of Quantum Shot Noise in Coupled Mesoscopic Systems}

\author{Masayuki Hashisaka, Yoshiaki Yamauchi, Shuji Nakamura, Shinya Kasai,\\
Teruo Ono, and Kensuke Kobayashi} 

\affiliation{Institute for Chemical Research, Kyoto University, Uij, Kyoto 611-0011, Japan}

\date{\today}

\begin{abstract}

We present a new scheme to detect the quantum shot noise in coupled mesoscopic systems. By applying the noise thermometry to the capacitively coupled quantum point contacts (QPCs) we prove that the noise temperature of one QPC is in perfect proportion to that of the other QPC which is driven to non-equilibrium to generate quantum shot noise. We also found an unexpected effect that the noise in the source QPC is remarkably suppressed possibly due to the cooling effect by the detector QPC. 

\end{abstract}

\pacs{73.23.Ad, 72.70.+m, 73.50.Lw, 73.50.Pz}

% 73.23.Ad    Ballistic transport
% 72.70.+m    Noise processes and phenomena  
% 73.50.Lw    Thermoelectric effects  
% 73.50.Pz    Photoconduction and photovoltaic effects 

\maketitle

Detection of the quantum-mechanical states in two-level systems (TLSs) is a key technique for the quantum information technology \cite{Nielsen2000,Awschalom2002}. The coupled mesoscopic device of a quantum dot (QD) and a quantum point contact (QPC) is one of the promising candidates of the qubit as an ideal combination of a TLS and a fast read-out scheme \cite{Awschalom2002,HansonRMP2007}. The most significant drawback of this technique is, however, the decoherence of the quantum states in the QD caused by the back-action from the QPC due to the shot noise \cite{AleinerPRL1997,LevinsonEL1997,BuksNature1998,KalishPRL2004}. The recent study on the non-local noise detection by using TLSs \cite{DeblockScience2003} illustrates the significance of the back-action in the QD-QPC coupled systems \cite{AguadoPRL2000,OnacPRL2006,GustavssonPRL2007}. In the experiments on the QD-QPC system \cite{OnacPRL2006,GustavssonPRL2007}, it is demonstrated that the high frequency noise is converted to the DC signal by using the energy level separations in QD, resulting in the very high sensitivity in the narrow bandwidth as a non-local noise detection scheme. In order to develop less disturbing scheme for the quantum state read-out, it is necessary to evaluate and minimize the total disturbance due to the shot noise, which has not been performed so far.

In this Letter, we report the on-chip bolometric noise detection scheme; the shot noise generated in one QPC is detected by using the other QPC in the capacitively-coupled double QPC (DQPC) system. The capacitors are designed to electrically decouple the two QPCs at low frequencies and also designed to transfer the high-frequency noise from one to the other. As the QPC constitutes the energy continuum, the electrical fluctuation in the source QPC can be detected for large frequency bandwidth, enabling us to evaluate the total disturbance by using the precise noise thermometry \cite{SpietzScience2003,GiazottoRPM2006} in the detector QPC, which is at the heart of our experiment. We also found an unexpected effect that the source QPC seems to be ``cooled" by the detector QPC, which would be potentially useful for the less disturbing charge-detection scheme.

\begin{figure}[b]
\includegraphics[width=0.98\linewidth]{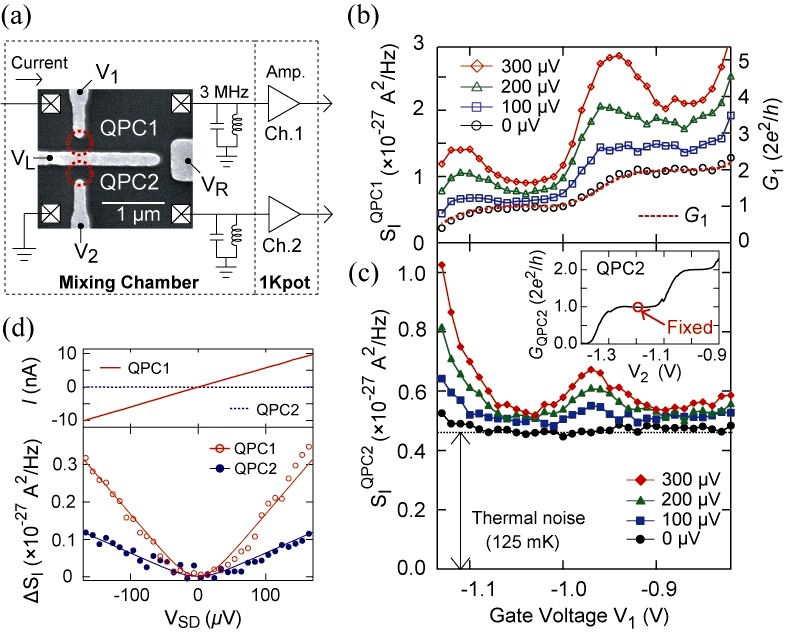}
\caption{\label{fig1} (a) Schematic diagram of the measurement setup with the scanning electron microscope image of the sample fabricated on 2DEG. (b) Current noises in QPC1 ($S_{I}^{\rm{QPC1}}$) measured at several source-drain bias voltages ($V_{\rm{SD}} =$ 0, 100, 200, and 300~$\mu$V) are shown (in the left axis) with the conductance of QPC1 (in the right axis) as a function of $V_{1}$. (c) Current noises in QPC2 ($S_{I}^{\rm{QPC2}}$) as a function of the gate voltage $V_{1}$ at several $V_{\rm{SD}}$ for QPC1. The inset shows the conductance of QPC2 as a function of $V_{2}$. (d) DC current and excess current noise ($\Delta S_{I}$: thermal noise is subtracted) in QPC1 and QPC2 as a function of $V_{\rm{SD}}$ measured at $V_{1} =$ -1.10~V. }
\end{figure}
Figure~1(a) shows the scanning electron microscope image of the DQPC device fabricated on the GaAs/AlGaAs two-dimensional electron gas (2DEG: electron density $= 2.3\times10^{11}~\rm{cm}^{-2}$ and mobility $= 1.1\times10^{6}~\rm{cm}^{2}/$Vs) and the measurement setup. DQPC was realized by applying the negative voltage to the side gate electrodes ($V_{1}$ and $V_{2}$) with applying fixed negative voltages to the center gate electrodes ($V_{L}=-1.4$ V and $V_{R}=-1.8$~V). We applied finite DC source-drain voltages ($V_{\rm{SD}}$) to QPC1 to produce shot noise while QPC2 was kept unbiased. The characteristic of the conductance of QPC2 ($G_{2}$) as a function of $V_{2}$ is shown in the inset of Fig.~1(c). Throughout our experiment, $G_{2}$ was fixed to $2e^{2}/h \cong (12.9~\rm{k}\Omega)^{-1}$, namely at the center of the first conductance plateau. The current noises in QPC1 and QPC2 were individually and simultaneously obtained by measuring the voltage fluctuation at the resonant circuits at 3.0 MHz through the two signal amplifying lines \cite{RobinsonRSI2004,DiCarloRSI2006,HashisakaPSS2008,HashisakaJPC2008}. 
The noise measurements were performed by capturing the time domain data and converting the data to spectral density signal by the two-channel digitizer. The experiment was performed in the dilution refrigerator whose base temperature is 45~mK and the electron temperature ($T_{e}$) in the equilibrium states were estimated to be 125~mK by measuring the thermal noise of the QPC. We applied a slight magnetic filed (0.2~T) perpendicular to the 2DEG as performed in Ref.~\cite{DiCarloRSI2006,KumarPRL1996}.

Now we show how the bolometric detection works in this system. Figures~1(b) and 1(c) show the typical results of the noise measurements obtained at different $V_{\rm{SD}}$ for QPC1 and QPC2, respectively. The conductance of QPC1 ($G_{1}$) measured simultaneously with the noise measurements is also shown in Fig.~1(b) (the right axis). For $V_{\rm{SD}} = 0~\mu$V, the behavior of the current noise in QPC1 exactly traces that of $G_{1}$ as a function of $V_{1}$ as expected for the thermal noise at $T_{e}$. In general, the shot noise occurs at finite $V_{\rm{SD}}$ due to the partition process of electrons at QPC \cite{KumarPRL1996,BlanterPR2000,ReznikovPRL1995}. In the low frequency limit, the power spectral density of the current noise ($S_{I}$) is given as follows;
\begin{alignat}{1}
\nonumber S_{I}(V_{\rm{SD}})&=4k_{B}T_{e}G\\
&+2FG\biggl( eV_{\rm{SD}}\coth \biggl( \frac{eV_{\rm{SD}}}{2k_{B}T_{e}}\biggr) -2k_{B}T_{e} \biggr),
\label{eq1}
\end{alignat}
where $G$ is the conductance of the QPC, $F$ is the Fano factor expressed as $F=\sum_{n}T_{n}(1-T_{n})/\sum_{n}T_{n}$ ($T_{n}$ is the transmission of the $n$-th channel in QPC), $k_{\rm{B}}$ is the Boltzmann constant, and $-e$ is the charge of electron. The measured noise in QPC1 ($S_{I}^{\rm{QPC1}}$) shown in Fig.~1(b) qualitatively agrees with the above theory; $S_{I}^{\rm{QPC1}}$ peaks as a function of $V_{1}$ when $G_{1}$ is around half-integer of $2e^{2}/h$, while it is suppressed at the center of the conductance plateaus (quantitative analysis will be discussed later in Fig.~4(b)). On the other hand, although QPC2 is kept unbiased, the noise in QPC2 ($S_{I}^{\rm{QPC2}}$) increases according to the increase of $S_{I}^{\rm{QPC1}}$ (Figs.~1(c) and 1(d)). Note that $S_{I}^{\rm{QPC2}}$ is insensitive to both the thermal noise and Joule 
heating of QPC1 but is only sensitive to the shot noise in QPC1. The thermal noise of QPC2 ($S_{I}^{\rm{QPC2}}$ at $V_{\rm{SD}} = 0~\mu$V) remains unaffected when $G_{1}$ changes as a function of $V_{1}$ (see the black curve in Fig.~1(c)), which clearly indicates that two QPCs are electrically decoupled at 3.0~MHz. $S_{I}^{\rm{QPC2}}$ also behaves independently of the Joule heating ($= V_{\rm{SD}}^{2}\times G_{1}$) expected for QPC1. In addition, Fig.~1(d) tells that no finite DC current is produced at QPC2 by the DC current flowing across QPC1, which excludes the possibility of the Coulomb-mediated or phonon-mediated drag effects \cite{DebrayJPCS2001,YamamotoScience2006,KhrapaiPRL2007}. We confirmed that no appreciable cross-correlated signal between QPC1 and QPC2 was detected at 3.0~MHz. The above findings clearly prove that QPC2 exactly works as a non-local shot noise detector, while two QPCs are electrically decoupled at DC and 3.0~MHz.

\begin{figure}[b]
\includegraphics[width=0.7\linewidth]{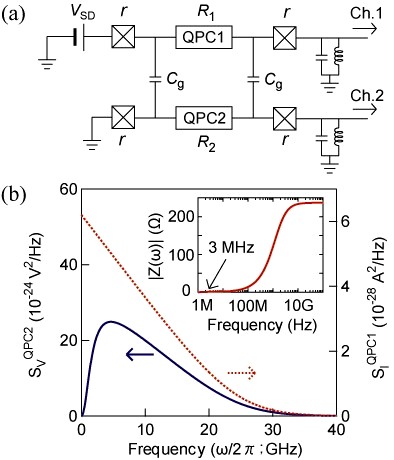}
\caption{\label{fig2} (a) Electrical circuit model of the present setup. The capacitances of the center gates ($C_{\rm{g}}$) were designed to be $C_{\rm{g}} \sim$ 200~fF. The typical resistance of the Ohmic contacts ($r$) including the lead resistances were 250~$\Omega$. (b) Current fluctuation at QPC1 (dotted red line) and the voltage fluctuation at QPC2 (blue line) for the finite frequency calculated for $G_{1} = e^{2}/h$ ($F =$ 0.5), $V_{\rm{SD}} =$ 100~$\mu$V, $G_{2} = 2e^{2}/h$, and $T_{e} =$ 125~mK. The inset shows the frequency dependence of the absolute value of the trnasimpedance Z($\omega$). The thermal noise is not taken into account in this plot. }
\end{figure}
We explain the mechanism of this non-local shot noise detection as follows; (1) At finite $V_{\rm{SD}}$, the high frequency current fluctuation occurs due to the shot noise in QPC1, (2) the high frequency fluctuation is conducted from QPC1 to QPC2 via the capacitance of the center gate electrodes ($C_{\rm{g}}$), and (3) the transferred fluctuation in QPC2 are energetically relaxed in 2DEG of QPC2 to increase $S_{I}^{\rm{QPC2}}$. In the first process, the spectral density of the high-frequency quantum shot noise for the emission side ($\omega > 0$) in the low-temperature limit ($\hbar \omega \gg k_{B}T_{e}$) is represented as \cite{AguadoPRL2000};
\begin{alignat}{1}
\nonumber S_{I}^{\rm{shot}}(\omega)=&\frac{4e^{2}}{h}\sum_{n} T_{n}(1-T_{n})\\
&\times \frac{eV_{\rm{SD}}-\hbar \omega}{1-\exp \Bigr(-(eV_{\rm{SD}}-\hbar \omega )/k_{\rm{B}}T_{e}\Bigl)}.
\label{eq2}
\end{alignat}
In the second process, the transport of the noise in the DQPC system is explained by using the electric circuit model shown in Fig.~2(a). QPC1 and QPC2 are capacitively coupled through $C_{\rm{g}}$ which was geometrically designed to be about 200 fF. $C_{\rm{g}}$ is sufficiently small that QPC1 and QPC2 are electrically disconnected for the low frequency (DC~-~100 MHz), while $C_{\rm{g}}$ can transfer signals above $\sim$ 1~GHz from QPC1 to QPC2. In fact, the voltage fluctuation in QPC2 ($S_{I}^{\rm{QPC2}}$) caused by the current fluctuation at QPC1 \cite{AguadoPRL2000,DeblockScience2003,OnacPRL2006} is given as $S_{I}^{\rm{QPC2}}(\omega) = |Z (\omega)|^{2} S_{I}^{\rm{QPC1}}(\omega)$, where the transimpedance $Z(\omega)=2iR_{1}R_{2}r^{2}\omega C_{\rm{g}}( (R_{1}+2r)(R_{2}+2r)+2i(R_{1}R_{2}+R_{1}r+R_{2}r)r\omega C_{\rm{g}})^{-1}$.
Here $R_{1}$, $R_{2}$, and $r$ are the resistances of QPC1, QPC2, and the four relevant Ohmic contacts. The simulated $S_{I}^{\rm{QPC1}}(\omega)$ and $S_{I}^{\rm{QPC2}}(\omega)$ in the equivalent circuit (Fig.~2(a)) with the typical values are presented in Fig.~2(b) with $|Z(\omega)|$ shown in the inset, meaning that QPC2 can detect the GHz fluctuations at QPC1.

\begin{figure}[t]
\includegraphics[width=0.98\linewidth]{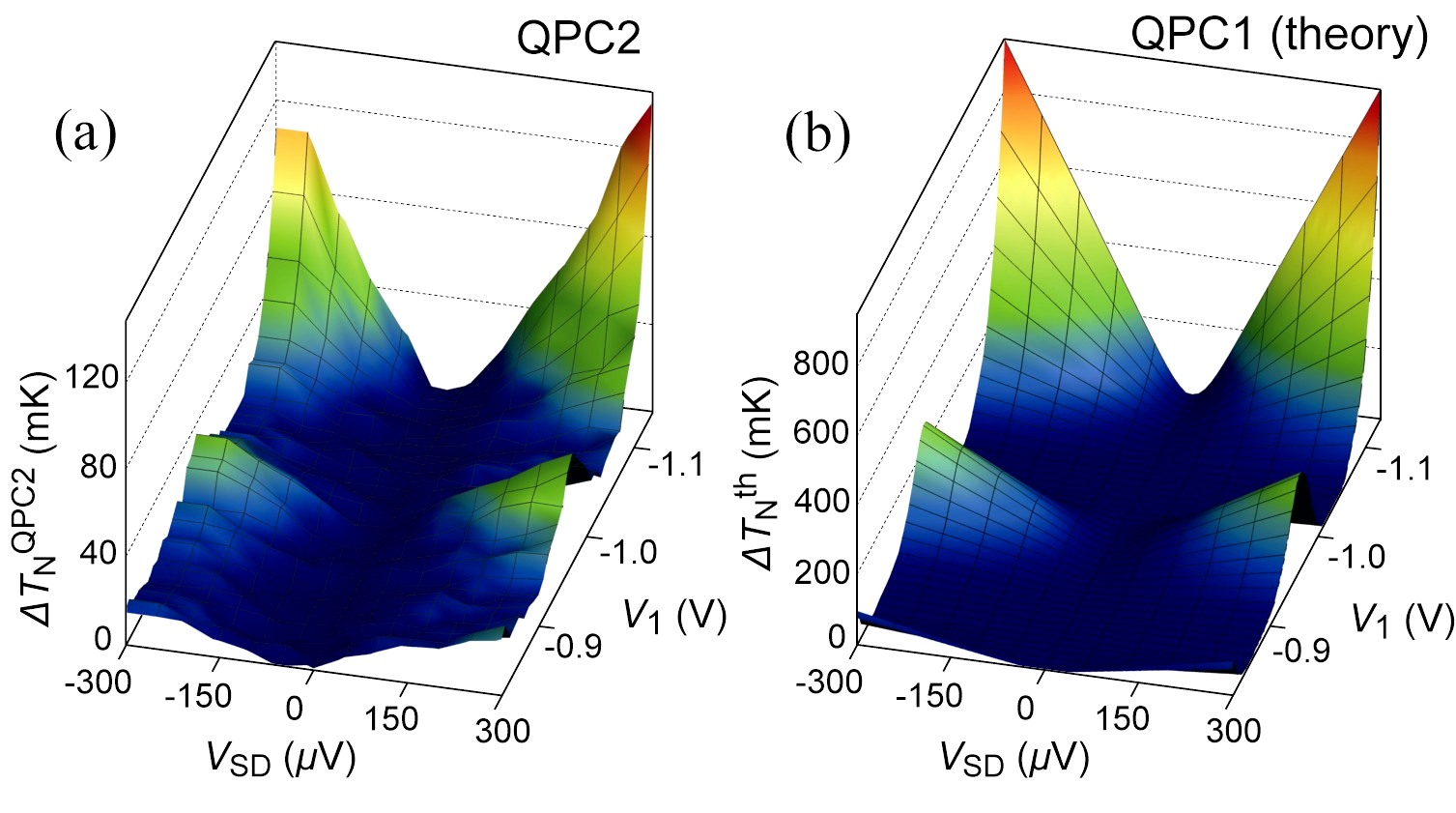}
\caption{\label{fig3} (a) 3D Image plot of the excess noise temperature measured for QPC2 ($\Delta T_{N}^{\rm{QPC2}}$). The horizontal axes are the gate voltage $V_{1}$ and the source-drain voltage $V_{\rm{SD}}$ applied to QPC1, respectively. (b) Corresponding 3D image plot of the excess noise temperature in QPC1 ($\Delta T_{N}^{\rm{th}}$) calculated based on the standard shot noise theory.}
\end{figure}
As the two QPCs are electrically decoupled at 3.0~MHz, the increase of the thermal noise in QPC2 at this frequency is beyond the circuit model and, therefore, we have to assume the presence of the rapid energy relaxation process of the transmitted GHz fluctuations to the lower frequencies. Such down-conversion of the high frequency noise is a direct result of the many-body correlation in 2DEG. As the GHz fluctuation dissipates in QPC2 and is absorbed by the thermal bath cooled by the dilution refrigerator, the energy continues to be transferred from QPC1 to QPC2. The energy-exchange process can be regarded as the absorption of energy from a high-temperature non-equilibrium system (QPC1) to a low-temperature system (QPC2).

\begin{figure}[t]
\includegraphics[width=0.70\linewidth]{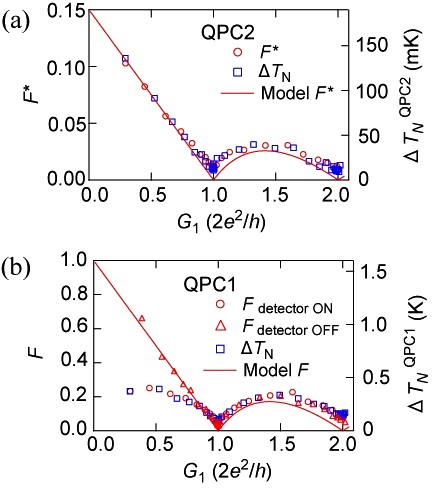}
\caption{\label{fig4} (a) Excess noise temperature at $V_{\rm{SD}} = 300~\mu$V (blue square: right axis) and the estimated rescaled Fano factor for QPC2 (red circle: left axis) as a function of the conductance of QPC1. The behavior was well fitted with the rescaled Fano factor (red line) by using Eqn.~(3). (b) As a function of the conductance of QPC1 are plotted the excess noise temperature at $V_{\rm{SD}} = 300~\mu$V (blue square: right axis) and the estimated Fano factor (red circle: left axis) for QPC1 with QPC2 fixed at the 1st conductance plateau, namely with the detector ``switched on". We also plot the Fano factor for QPC1 (red triangle: left axis) when the resistance of QPC2 is set to $\sim$ 0~$\Omega$, namely with the detector ``switched off".}
\end{figure}
To be more quantitative, we compare the excess noise temperature of QPC2 ($\Delta T_{N}^{\rm{QPC2}}$) defined by $S_{I}^{\rm{QPC2}} = 4k_{\rm{B}}(T_{e}+\Delta T_{N}^{\rm{QPC2}})G_{2}$ with the theoretical excess noise temperature of QPC1 ($\Delta T_{N}^{\rm{th}}$) derived from Eq.~(1) in Figs.~3(a) and 3(b). Clearly, $\Delta T_{N}^{\rm{QPC2}}$ is proportional to $\Delta T_{N}^{\rm{th}}$, indicating that QPC1 behaves to QPC2 as if it is the heater with the excess temperature of $\Delta T_{N}^{\rm{th}}$. Now, we estimate the energy transferred from QPC1 to QPC2 as well as the efficiency of QPC2 as a bolometer. In Fig.~4(a) where the measured $\Delta T_{N}^{\rm{QPC2}}$ at $V_{\rm{SD}} = 300~\mu$V is shown in the right axis, we found that $\Delta T_{N}^{\rm{QPC2}}$ is perfectly proportional to the Fano factor expected for QPC1. We define the efficiency ($A$) as a scaling factor of $F~(F^{*} = A\times F)$, where the rescaled Fano factor $F^{*}$ is obtained from the fitting for the measured current fluctuation (see Fig.~1(d)) by the following fitting function;
\begin{alignat}{1}
\nonumber S_{I}^{\rm{QPC2}}&(V_{\rm{SD}})=4k_{B}T_{e}G_{2}\\
&+2F^{*}G_{2}\biggl( eV_{\rm{SD}}\coth \biggl( \frac{eV_{\rm{SD}}}{2k_{B}T_{e}}\biggr) -2k_{B}T_{e} \biggr).
\label{eq4}
\end{alignat}
The obtained efficiency $A =$ 0.15 (see Fig.~4(a)) implies that at least 15~\% 
of the energy generated by the shot noise in the biased QPC1 is continuously absorbed by QPC2 \cite{NonLocal}. This estimation gives the lower limit of the transferred energy which depends on the thermal conductivity in QPC2, since QPC2 is connected to the thermal bath of the dilution refrigerator; if the thermal conductivity is lower, $A$ becomes larger and vice versa. The emission power from the source to the detector is also estimated to be $\int d\omega |Z(\omega)|S_{I}^{\rm{QPC1}}(\omega) \cong 1.7$~fW for the same parameters as in Fig.~2(b), which gives the typical sensitivity of our bolometer \cite{Overestimate}.

So far we have focused ourselves in establishing that the quantum noise of QPC1 ``heats" QPC2. What happens in our noise source QPC1? Surprisingly, we found an indication that the inverse process might be present; QPC2 ``cools" QPC1 by absorbing the high frequency shot noise component presumably because QPC2 works as a dissipative environment for QPC1 \cite{NaikNature2006}. In Fig.~4(b), we plot the observed excess noise temperature and the Fano factor in QPC1 when the QPC2 was fixed at the first conductance plateau (see the inset of Fig.~1(c)). Experimentally, the ``cooling" effect appears at the low conductance region ($G_{1} < 2e^{2}/h$), where the excess noise temperature of QPC1 defined as above ($\Delta T_{N}^{\rm{QPC1}}$) is significantly lower than $\Delta T_{N}^{\rm{th}}$. In contrast, in Fig.~4(b), we superpose the Fano factor obtained when the resistance of QPC2 was set to $\sim$ 0 $\Omega$ ($V_{2} =$ 0~V), which gives the Fano factor as the theory expects. This clearly tells that the ``cooling" effect occurs only when QPC2 works as a bolometer, indicating that the decrease of Fano factor in QPC1 is a back-action from the bolometric detection. The suppression of the noise temperature, namely the Fano factor suppression, in QPC1 becomes striking with the decrease of $G_{1}$. However, the effect is little, if any, at $G_{1} > 2e^{2}/h$, while the energy absorption by QPC2 occurs as QPC2 detects the correct shot noise (Fig.~4(a)). Insufficiency of the resolution of our measurement system or other essential energy transfer mechanisms may prevent us to detect the small decrease at the higher conductance region. Further investigation of the ``cooling" effect would be useful in creating less disturbing and more accurate charge read-out scheme in the QD-QPC system, where the shot noise would not give the principle limit of accuracy \cite{VandersypenAPL2004}.

To conclude, we demonstrate the mesoscopic bolometry in the capacitively-coupled double QPCs, where the shot noise at one QPC is detected by the increase of the noise temperature of the other QPC. While the shot noise is described in the scattering theory in quantum mechanics \cite{BlanterPR2000}, it can heat materials by being dissipated when the system is connected to the other with the appropriate impedance. We have also found that the inverse process, namely the cooling effect might be present, which may be useful in creating less disturbing charge read-out scheme in the QD-QPC system. Moreover, the simple detection scheme of the present noise bolometry makes it attractive for diverse ultra-precise measurements such as in the bolometric photon counting and the advanced metrology.

%acknowledgement
We appreciate fruitful discussion with R. Leturcq, Y. Chung, S. Gustavsson, C. Strunk, P. Roche, T. Martin, Y. Utsumi, T. Fujii and K. Matsuda. This work is supported by KAKENHI, Yamada Science Foundation, and Matsuo Science Foundation. MH thanks for the financial support by SCAT and JSPS Research Fellowships for Young Scientists. 

%%reference


\begin{thebibliography}{}
\bibitem{Overestimate} The value is overestimated as the transimpedance should be lowered from the estimation 
in the simple circuit model due to the stray capacitance around the DQPC sample \cite{DeblockScience2003}.
\end{thebibliography}
\end{document}